\documentclass[a4paper]{jpconf}

\usepackage{graphicx}
\usepackage{epsfig,eufrak}

\begin{document}

\title{Non-equilibrium critical dynamics in disordered ferromagnets}

\author{Gr{\'e}gory Schehr$^1$, Raja Paul$^2$}

\address{$^1$ Theoretische Physik, Universit\"at des Saarlandes,
66041 Saarbr\"ucken, Germany }
\address{$^2$ BIOMS, IWR, Ruprecht-Karls-University Heidelberg, 69120
  Heidelberg }

\ead{schehr@lusi.uni-sb.de}

\begin{abstract}
We discuss some aspects of the non-equilibrium relaxational
dynamics which occur after a quench at a
disordered critical 
point. In particular, we focus on the violation of the fluctuation
dissipation theorem for local as well as non-local observables and on
persistence properties.   
\end{abstract}

Although critical dynamics has been a subject of study for many years
\cite{hohenberg_review}, 
it was rather recently recognized \cite{cugliandolo_fdr_pure,
  godreche_crit_ferro_review}that, although simpler to study than 
glasses, they display interesting non-equilibrium
features such as aging or violation of the Fluctuation Dissipation
Theorem (FDT), commonly observed in more complex disordered or  
glassy phases \cite{cugliandolo_leshouches}. In the same context, 
relaxational dynamics at pure critical point has been the subject of
numerous analytical as well as numerical recent studies
\cite{calabrese_review_fdr}. Interestingly, it has been proposed
\cite{godreche_crit_ferro_review} 
that a non trivial Fluctuation   
Dissipation Ratio (FDR) $X$, originally introduced in the 
Mean Field approach to glassy systems, which  
generalizes the FDT to non
equilibrium situations, is a new {\it universal} quantity
associated to these critical points. As such, it has been computed using the
powerful tools of RG, {\it e.g.} for pure $O(N)$ model at
criticality in the vicinity of the upper critical 
dimension $d_{\mathrm{uc}} = 4$ and for various dynamics
\cite{calabrese_review_fdr}.   

On the other hand, critical dynamics
display interesting 'global persistence' properties.  Indeed, 
it has been shown~\cite{majumdar_critical} that the 
probability ${P}_{{c}}(t)$ that the global magnetization
$M$ has not changed sign in the time interval $t$ following a quench
from a random 
initial configuration, 
decays algebraically at large time ${P}_c(t) \sim
t^{-\theta_c}$. In this context, analytical progress is made
possible, thanks to the 
property that, in the thermodynamic limit, the global order parameter
remains Gaussian at all finite 
times $t$. Indeed, for a $d$-dimensional system of linear size $L$,
$M(t)$ is the sum 
of $L^d$ random 
variables which are correlated only over a {\it finite} correlation length
$\xi(t)$. Thus, in the thermodynamic limit $L/\xi(t) \gg 1$, the
Central Limit Theorem (CLT) asserts that $M(t)$ is a Gaussian
process, for which powerful tools have been developed to compute the
persistence
properties~\cite{satya_review,satya_clement_persist}. Remarkably,
under the {\it additional} assumption  
that $M$ is a Markovian process, $\theta_c$ can be related to the
other critical exponents via the scaling relation $\theta_c =\mu
\equiv (\lambda_c - d + 1 - \eta/2)z^{-1}$, with $z$ and $\lambda_c$ the
dynamical and 
autocorrelation~\cite{janssen_rg, huse_lambda} exponent
respectively, and $\eta$ the (static) Fisher exponent. Nevertheless,
as argued in  
Ref.~\cite{majumdar_critical}, $M$ is in general non Markovian and
thus $\theta_c$ is a {\it new} exponent associated to critical
dynamics. For the non conserved critical dynamics of
pure ${O}(N)$ model, corrections to this scaling relation were
indeed found to occur at two-loops order~\cite{oerding_persist}, 
in rather good agreement with numerical simulations in dimensions
$d=2,~3$~\cite{Schulke97}.

Characterizing the effects of quenched disorder on
critical dynamics is a complicated task and indeed  
the question of how quenched randomness modifies these 
properties has been much less studied. In these notes, 
we address these
questions on one prototype of such 
disordered ferromagnets, the randomly diluted Ising model: 
\begin{eqnarray}\label{def_diluted}
H = - \sum_{\langle i j \rangle} \rho_i \rho_j s_i s_j
\label{eq_Hamil}
\end{eqnarray}  
where $s_i$ are Ising spins on a $d$-dimensional hypercubic lattice
and $\rho_i = 1$ with probability $p$ and
$0$ with probability $1-p$. For the experimentally relevant case of
dimension $d=3$ \cite{belanger_exp}, for which the specific heat
exponent of the pure 
model is positive, the disorder is expected, according to Harris
criterion\cite{harris_criterion}, to modify the universality class of the
transition. For $1-p \ll 1$,  the
large scale properties of (\ref{def_diluted}) at criticality
are then described by the
following ${O}(1)$ model with a random mass term, the so-called Random
Ising Model (RIM)\cite{folk_rim_review}: 
\begin{equation}\label{H_rim}
H^{\psi}[\varphi] = \int d^d x \left[ \frac{1}{2}(\nabla \varphi)^2 +
 \frac{1}{2} [r_0 + \psi(x)] \varphi^2 + \frac{g_0}{4!}
 \varphi^4 \right]  
\end{equation} 
where $\varphi \equiv \varphi(x)$ and  
$\psi(x)$ is a Gaussian random variable $\overline{\psi(x)
  \psi(x')} = \Delta \delta^d(x-x')$ and $r_0$, the bare mass, is
  adjusted so that the renormalized one is zero.

\section{Non equilibrium dynamics of one time quantity : initial slip
  exponent.}

\begin{figure}[t]
\includegraphics[width=18pc]{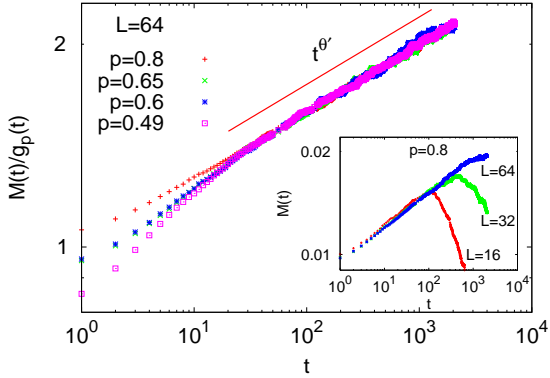}\hspace{2pc}
\begin{minipage}[b]{18pc}\caption{Rescaled global magnetization $M(t)/g_p(t)$ as a function $t$ for $p=$ 0.49, 0.6, 0.65 and 0.8 in the log-log scale. The linear system size is $L=64$, the initial magnetization is $M_0 = 0.01$ and the measured exponent $\theta'=0.1$. {\bf Inset:} Global magnetization $M(t)$ as a function of time $t$ for $p=0.8$ and $L=16, 32$ and $64$.}
\end{minipage}
\label{fig_initial_slip}
\end{figure}

We first focus on one-time quantity, the global magnetization $M(t)$
\begin{eqnarray}
M(t) = \frac{1}{N_{\mathrm{occ}}} \sum_i \rho_i s_i(t) = \frac{1}{L^d}\int_x
\varphi(x,t), \label{def_mag}
\end{eqnarray}
which already carries the signatures of a non-equilibrium situation. 
To observe them, we study numerically the time evolution of $M(t)$ 
when the system is quenched from an initial configuration with short
range correlations but a finite, however small, magnetization
$M_0$. The initial stage 
of the dynamics is characterized by an 
increase of the global magnetization, described by a universal power
law~\cite{janssen_rg}
\begin{equation}
M(t) \sim M_0 t^{\theta'} \label{eq_ini_slip}
\end{equation}
At larger times, $t\gg t_0$, critical fluctuations set in the system
and cause the decrease of $M(t)$ to zero as $M(t) \sim
t^{-(d-2+\eta)/(2 z)}$. 
In our simulations the system 
is initially prepared in a random initial configuration with mean
magnetization $M_0 = 0.01$. At each time step, one
site is randomly chosen and the move  
$s_i \to - s_i$ is accepted or
rejected according to Metropolis rule. One time-unit corresponds to
$L^3$ such time steps. In
all subsequent times we measure $M(t)$ (\ref{def_mag})
for linear system sizes $L=$ 8, 16, 32 and 64. Finally data are
averaged over $8 \times 10^5$ samples for $L=8$ to $10^4$ samples for
$L=64$. In the 
inset of Fig.~\ref{fig_initial_slip}, we show a plot of $M(t)$ for
$p=0.8$ and different system sizes. One sees clearly that $M(t)$ is
increasing until a time $t_0$, which is an increasing function of $L$ 
for the sizes
considered here~\footnote[1]{In the thermodynamic limit, one expects
  that it reaches the asymptotic value 
$t_0 \sim M_0^{-1/(\theta'+(d-2+\eta)/(2 z))}$~\cite{janssen_rg}.} and
compatible with the scaling $t_0 
\sim L^z$, and then decreases to
zero (although the 
aforementioned scaling for $t \gg t_0$ is not clearly seen here). 
By computing $M(t)$ for different values of $p=0.499, 0.6, 0.65$
and $0.8$, we observe corrections to scaling, which are know to be
strong in this model~\cite{heuer_tc,parisi_simu_rim}. Following
Ref.~\cite{parisi_simu_rim, us_pre},  
we take them into
account as $M(t) = t^{\theta'} g_p(t)$ with
$g_p(t)=A'(p)[1+B'(p)t^{-b}]$, where $b=0.23(2)$ has been determined
previously \cite{parisi_simu_rim, us_pre}, $A'(p),~B'(p)$ being
fitting parameters.

As shown in
Fig.~\ref{fig_initial_slip}, one obtains a reasonably good data
collapse of $M(t)/g_p(t)$ 
vs.~$t$ for the different values of $p$. After a microscopic time
scale, one observes a universal
power law increase (\ref{eq_ini_slip}), from which we get the estimate
\begin{eqnarray}
\theta'=0.10(2), \label{initial_slip_num}
\end{eqnarray}
which is in good agreement with a previous 
two-loops estimate $\theta'_{\mathrm{2-loops}} = 0.0868$
\cite{oerding_randommass_theta_twoloops}.

\section{Two time quantities : Aging and Violation of
  FDT.}

Although non-equilibrium effects can already be observed in one time
quantity such as the global magnetization $M(t)$, it is also interesting to study two-times, $t,t_w$
functions, which we now focus on. 

\subsection{Analytical approach in $d=4-\epsilon$.}

We study the relaxational dynamics of the Randomly diluted Ising Model
in dimension $d=4-\epsilon$ described by a Langevin equation:
\begin{eqnarray}\label{def_Langevin}
\eta \frac{\partial}{\partial t} \varphi(x,t) = - \frac{\delta
 H^{\psi}[\varphi]}{\delta \varphi(x,t)} + \zeta(x,t)
\end{eqnarray}
where $\langle \zeta(x,t) \rangle = 0$ and
$\langle \zeta(x,t) \zeta(x',t') \rangle = 2\eta T \delta(x-x')
\delta(t-t') $ is the thermal noise and $\eta$ the friction
coefficient. At initial time $t_i = 0$, the system is in a random
initial configuration with {\it zero} magnetization $m_0=0$
distributed according to 
a Gaussian with short range correlations 
\begin{eqnarray}\label{ini_cond}
[\varphi(x,t=0)\varphi(x',t=0)]_i = \tau_0^{-1} \delta^{d}(x-x')
\end{eqnarray}
Notice that it has been shown that $\tau_0^{-1}$ is irrelevant (in the RG sense) in the large time regime studied here
\cite{janssen_rg}. We focus on the correlation ${\cal
C}^q_{tt_w}$ in Fourier space and the autocorrelation $C_{tt_w}$  
\begin{eqnarray}\label{def_C}
{\cal{C}}^q_{tt_w} =\overline{\langle {\varphi}(q,t)
      {\varphi}(-q,t_w)  \rangle} \quad, \quad  C_{tt_w} =
      \overline{\langle {\varphi}(x,t) 
      {\varphi}(x,t_w)  \rangle}
\end{eqnarray}
and the response ${\cal R}^q_{tt_w}$ to a small external
field ${f}(-q,t_w)$ as well as on the local response function
${R}^{}_{tt_w}$ respectively defined, for $t>t_w$   
\begin{eqnarray}\label{def_R}
{\cal R}^q_{tt_w} = \overline{\frac{\delta \langle
  {\varphi}(q,t) \rangle}{\delta 
  {f}(-q,t_w)} } \quad, \quad {R}^{}_{tt_w} 
 = \overline{\frac{\delta \langle {\varphi}(x,t) \rangle}{\delta 
  {f}(x,t_w)} },
\end{eqnarray}
where $\overline{..}$ and $\langle .. \rangle $
denote averages over disorder and 
thermal fluctuations respectively. In a previous work \cite{us_pre}, using the
exact renormalization group equation for the dynamical effective
action, we have computed these functions 
${\cal R}^q_{tt_w}, {\cal C}^q_{tt_w}$ up to one-loop order:
\begin{equation}
{\cal R}^{ {q}}_{ {t} {t_w}} =
 {q}^{-2+z+\eta}\left( \frac{ {t}}{ {t_w}} \right)^{\theta}
F_R^{eq}( {q}^z( {t}- {t_w})) \quad, \quad {\cal
 C}^{{q}}_{{t}{t_w}} = T_c 
{q}^{-2+\eta}\left( \frac{{t}}{{t_w}} \right)^{\theta}
F_C({q}^z({t}-{t_w}),{t}/{t_w}) \quad,
\label{janssenscalingcorr} 
\end{equation} 
with (up to order ${\cal O}(\sqrt{\epsilon)}$) : 
$\eta = 0$, $z = 2 + \sqrt{\frac{6\epsilon}{53}}$
\cite{grinstein_randommass_z_oneloop} and $\theta =
\frac{1}{2} \sqrt{\frac{6\epsilon}{53}}$
\cite{kissner_random_mass}, $\theta$ being related to the
autocorrelation exponent $\lambda_c$ through $\theta = 1 +
(d-2+\eta-\lambda_c)/z$. Notice that the scaling function $F_R^{eq}$
is  
a function of $q^z(t-t_w)$ only (an therefore the superscript $eq$),
which, although $z \neq 2$ is  
in agreement at this order with local
scale invariance arguments \cite{henkel_lsi}. In Ref. \cite{us_pre}, we have 
computed analytically the scaling functions $F_{{\cal R},{\cal C}}$ at
one-loop. And we will see later that some interesting information can
be extracted from them (see in particular section 3). Here we only
mention their asymptotic behaviors which may be relevant for our
study. Defining $v = q^z(t-t_w)$ and $u=t/t_w$
a first interesting scaling regime corresponds to $u \gg 1$, keeping
$v$ fixed. In that regime, one obtains an interesting relation
\begin{eqnarray}
&&F_C(v,u) = \frac{1}{u} F_{C,\infty}(v) + {\cal O}(u^{-2})
  \label{1_o_u} \\
&&F_{C\infty}(v) = A_{C \infty} v F_R^{{eq}}(v) \quad, \quad A_{C
  \infty} = 2 
  + 2\sqrt{\frac{6\epsilon}{53}} \label{relation}
\end{eqnarray}
The first relation (\ref{1_o_u}) is expected from general RG arguments
\cite{janssen_rg}, and it has also been checked for pure $O(N)$ models
\cite{calabrese_on_oneloop}. The second one (\ref{relation}) is not
predicted by such arguments, and it plays a crucial role in the
computation of the FDR. We notice that it has also been obtained in
the glassy phase of disordered elastic systems in $d=2$
\cite{schehr_co} as
well as in the relaxational dynamics near the depinning transition
\cite{schehr_depinning}. Whether this
relation holds for the present model at two-loops remains an open question.   

An other interesting asymptotic behavior, relevant when we want
extract information on local quantity from (\ref{janssenscalingcorr}),   
corresponds to $v \gg 1$, keeping $u$ fixed. In that regime, 
$F_R^{eq}(v)$ (as well as $F_C(v,u)$)
decays algebraically. In particular 
\begin{eqnarray}
F_R^{eq}(v) \sim v^{-2} \quad, \quad v \gg 1, \label{power_decay}
\end{eqnarray}
which is in sharp contrast with pure critical systems, {\it e.g.} pure
$O(N)$ models, where the decay is actually exponential
\cite{calabrese_on_oneloop}.  

From the response and correlation, it is interesting to compute the FDR $X^q_{tt_w}$ defined as~\cite{cugliandolo_leshouches}:
\begin{equation}
\frac{1}{{X}^q_{tt_w}} = \frac{\partial_{t_w} {\cal
    C}^q_{tt_w}}{T {\cal R}^q_{tt_w}},
 \label{def_FDR}
\end{equation}
such that ${{X}^q_{tt_w}} = 1$ at equilibrium. From the scalings
obtained above (\ref{janssenscalingcorr}), one has directly $X^q_{tt_w} \equiv
F_{X}(q^z(t-t_w),t/t_w)$. In particular, in the large $t/t_w$ limit,
keeping $q^z(t-t_w)$ fixed, one has from (\ref{relation})
\begin{eqnarray}\label{Xq}
\lim_{u \to \infty} \left({X}^q_{tt_w}\right)^{-1} = 2 +
\sqrt{\frac{6\epsilon}{53}} + {\cal O}(\epsilon) 
\end{eqnarray}
{\it independently} of $v$, {\it i.e.} of (small) wave vector $q$, which
coincides of course with the asymptotic value for the $q=0$ mode
obtained in Ref. \cite{calabrese_fdr_randommass}.

It is also interesting to study the limiting value
of the FDR for different observables \cite{sollich,sollich_new}, and
in particular 
to focus on local quantities. Therefore we compute
the FDR associated 
to the autocorrelation $C_{ttw}$ (\ref{def_C}) and the local
response $R_{ttw}$ (\ref{def_R}) which can be written as
\cite{calabrese_on_oneloop}: 
\begin{eqnarray}
\frac{1}{{X}^{x=0}_{tt_w}} = \frac{\partial_{t_w} {C}_{tt_w}}{T
  {R}_{tt_w}} = \frac{\int_q {\cal
  R}^q_{tt_w}({X}^{q}_{tt_w})^{-1} }{\int_q {\cal R}^q_{tt_w}}
  \label{rel_fdr}
\end{eqnarray}
For pure critical systems, ${\cal R}^q_{tt_w}$ decays exponentially
for large $q^z t$. And thus, in the limit $t \gg 1$, the integral over
the Fourier modes $q$ in (\ref{rel_fdr}) is dominated by the $q=0$
mode. One thus expects from this heuristic argument
\cite{calabrese_on_oneloop} that, 
\begin{eqnarray}
\lim_{t_w \to \infty} \lim_{t \to
  \infty} X^{x=0}_{tt_w} =  \lim_{t_w \to \infty} \lim_{t \to
  \infty} X^{q=0}_{tt_w} \label{rel_x_inf}
\end{eqnarray}
We have, however, seen that for the dilute Ising model, the response
function ${\cal R}^q_{tt_w}$ decays algebraically for $q^z t \gg 1$
(\ref{power_decay}) so that this heuristic argument does not
hold. And therefore, this relation (\ref{rel_x_inf}), if true at
all, is far from trivial for the present model. To verify this, we need a
direct computation in real space.  

Having obtained the scaling functions associated to ${\cal
  C}^q_{tt_w}$ and ${\cal R}^q_{tt_w}$ for any Fourier mode $q$, 
we obtain $C_{tt_w}$ and $R_{tt_w}$ by Fourier transform \cite{us_pre}.   
Due to the algebraic large $q$ behavior obtained previously
  (\ref{power_decay}), one obtains logarithmic corrections to scaling
\begin{equation}\label{scal_auto_correl}
{R}^{}_{tt_w} = \frac{K_d}{2}
\frac{A^0_{\cal R} + A^1_{\cal R}
\ln{(t-t_w)}}{(t-t_w)^{1+(d-2+\eta)/z}}\left(\frac{t}{t_w}\right)^{\theta} 
\quad,
\quad {C}^{}_{tt_w} = K_d  
\frac{A^0_{C} + A^1_{C}
\ln{(t-t_w)}}{(t-t_w)^{(d-2+\eta)/z}}\left(\frac{t}{t_w}\right)^{\theta}
{\cal F}(t/t_w)  
\end{equation}
with ${\cal F}(u) = {1}/{(1+u)} + {\cal O}(\epsilon)$ and where
$A^{0,1}_{\cal R, \cal C}$ are non-universal amplitudes. From these
expressions (\ref{scal_auto_correl}) one can then compute the
local FDR under the form
\begin{eqnarray}\label{local_FDR}
&&({X}^{x=0}_{tt_w})^{-1} = {\cal F}_{X}(t/t_w) \\
&&{\cal F}_{X}(u) = 2 \frac{u^2+1}{(u+1)^2} +
  \sqrt{\frac{6\epsilon}{53}} \left(\frac{u-1}{u+1}\right)^2 + {\cal
    O}(\epsilon)  
\end{eqnarray}
where ${\cal F}_{X}(u)$ is a monotonic increasing function of $u$ : it
interpolates between $1$, in the
quasi-equilibrium regime for $u\to 1$, and its asymptotic value for $u
\to \infty$ given by
\begin{equation}
\lim_{t_w\to \infty} \lim_{t\to \infty} ({X}^{x=0}_{tt_w})^{-1} =
\lim_{t_w\to \infty}\lim_{t\to \infty} 
({X}^{q=0}_{tt_w})^{-1} = 2 + \sqrt{\frac{6\epsilon}{53}} + {\cal
  O}(\epsilon)  \label{expl_calc} 
\end{equation}
which shows explicitly, at order ${\cal O}(\sqrt{\epsilon})$ that the
asymptotic FDR for both the global 
and the local magnetization are indeed in the same
(\ref{rel_x_inf}).

\subsection{Monte Carlo study : Autocorrelation functions.}

\begin{figure}[h]
\begin{minipage}[b]{0.43\linewidth} 
\includegraphics[scale=0.27,angle=0]{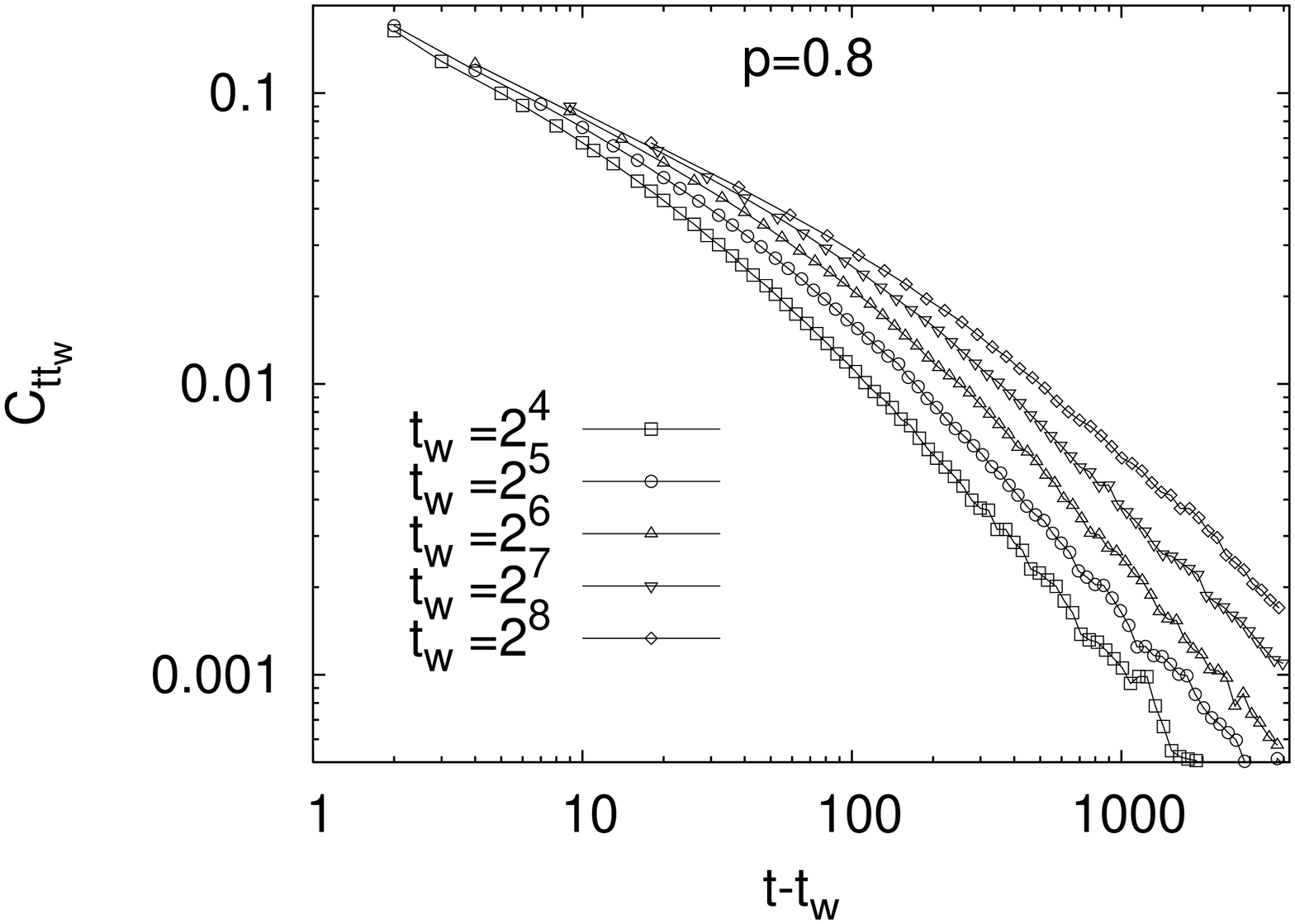}
\caption{Log-log plot of auto-correlation function ${\cal C}_{tt_w}$
 vs. $t-t_w$ for different waiting times. } 
\label{fig_correl1}
\end{minipage}\hspace{2pc}
\begin{minipage}[b]{0.43\linewidth}
\vspace*{0.2cm}
\includegraphics[scale=0.27,angle=0]{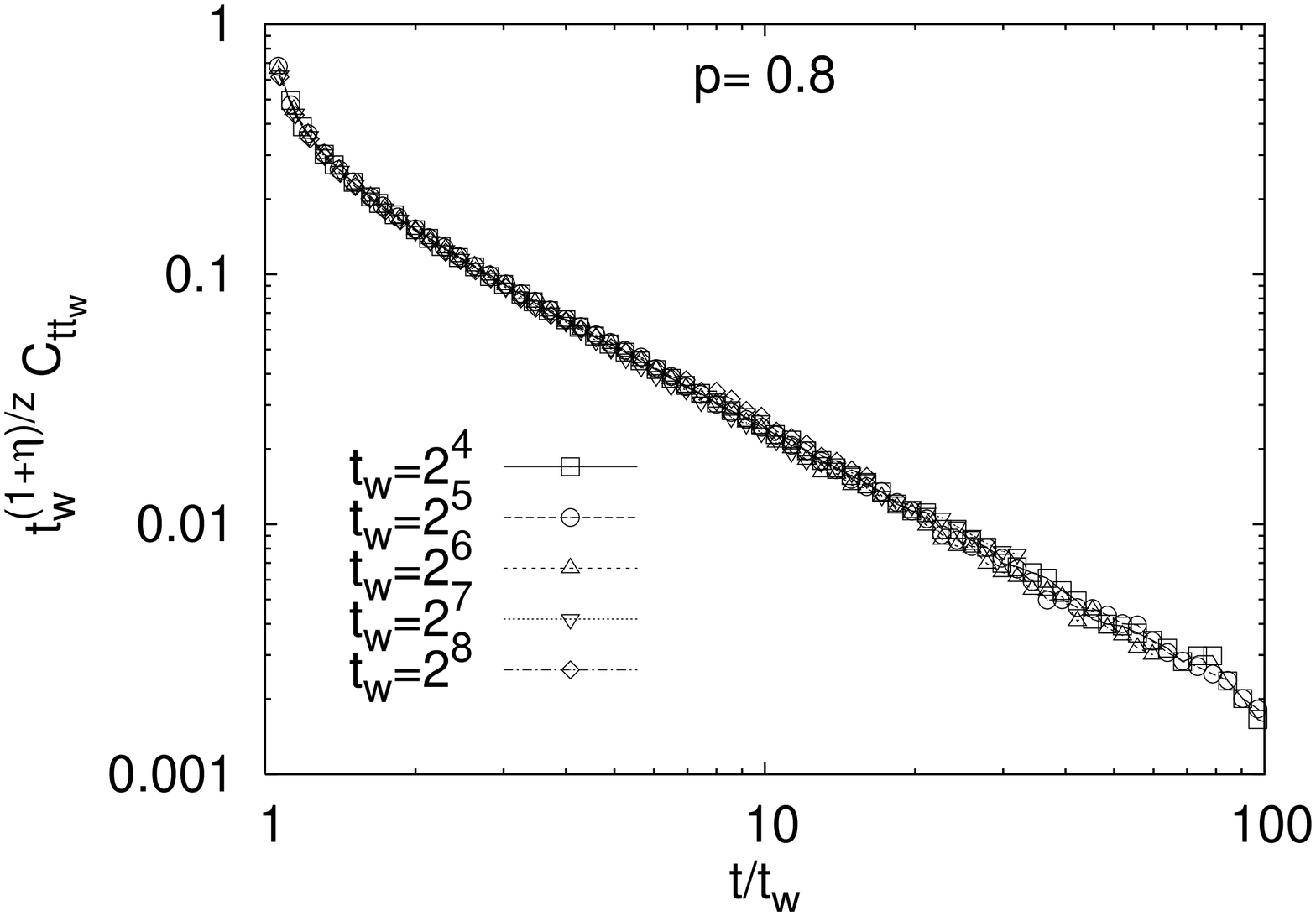}
\caption{Log-log plot of $t_w^{(1+\eta)/z}{\cal C}_{tt_w}$
 as function of $t/t_w$ for different waiting times.
}\label{fig_correl2}
\end{minipage}
\end{figure}

Let us next present results from our Monte Carlo simulations of the
 relaxational dynamics of the randomly diluted Ising model
 (\ref{eq_Hamil}) in dimension $d=3$, which, as previously (see
 section 1)  were done on 
 $L^ 3$ cubic lattices with periodic boundary
 conditions. The system is initially prepared in a random initial
configuration with zero magnetization, and at each time step, the $L^3$
 sites are sequentially updated according to Metropolis rule. Here we present
 our numerical data for $p=0.8$ (other values of $p$ 
are discussed in \cite{us_pre}). We compute the spin-spin
auto-correlation function defined as 
\begin{equation}\label{def_correl_num}
{C}_{tt_w} = {\frac {1}{L^3}}  \sum_{i} \overline{\langle
 s_i(t)~s_i(t_w) \rangle} 
\end{equation}
Fig.~\ref{fig_correl1} shows the auto-correlation function ${C}_{tt_w}$ as a
function of $t-t_w$ for different values of the waiting time 
$t_w = 2^4,~2^5,~2^6,~2^7$ and $2^8$ at $p=0.8$. 
One observes a clear
dependence on $t_w$, which indicates a non-equilibrium dynamical
regime. The scaling form obtained from the RG analysis
(\ref{scal_auto_correl}) suggest, discarding the logarithmic
correction, to plot $t_w^{(1+\eta)/z}{C}_{tt_w}$ as a function of
$t/t_w$. Taking the values $\eta = 0.0374$ from
Ref.~\cite{ballesteros_tc} and $z=2.62$ from Ref.~\cite{parisi_simu_rim},
we see in Fig.~\ref{fig_correl2} that, for $p=0.8$, 
one obtains a good collapse of the curves for
different $t_w$. As shown in Fig.~\ref{fig_correl2}, $t_w^{(1+\eta)/z}{C}_{tt_w}$ decays as a power
law, which allows to estimate the
value of the decaying exponent $\lambda_c/z$:
\begin{eqnarray}\label{exp_lambda}
\frac{\lambda_c}{z} = 1.05 \pm 0.03 
\end{eqnarray} 
We have checked \cite{us_pre}, taking carefully corrections to
scalings, that this decaying exponent (\ref{exp_lambda}) is actually
independent of the dilution factor $p$,  which supports universality
of the long-time non-equilibrium relaxation in this model. In
addition, using $z=2.6$, our numerical estimates of $\theta'$ and
$\lambda_c$ (\ref{exp_lambda}) are consistent with the scaling
relation $\lambda_c = d -z\theta'$.

\section{Persistence properties.}

Let us now focus on persistence properties of the dilute Ising model at criticality. Defining the global
magnetization $M(t)$ as in Eq. (\ref{def_mag})
we are interested in the disorder averaged probability
$\overline{{P}_c}(t)$ 
that the magnetization has 
not changed sign in the time interval $t$ following the quench.

\subsection{Analytical approach in $d=4-\epsilon$}

In a previous publication \cite{us_epl}, we have shown that the Exact
renormalization group equation allows 
to describe the time evolution of the magnetization by an
effective Gaussian process $\tilde M(t)$: 
\begin{eqnarray}
&&\partial_t \tilde M(t) + \sigma(t) \tilde M(t) = -\int_{0}^{t} dt_1
\Sigma_{tt_1} \tilde M(t_1) + {\tilde \zeta}(t) \label{Eff_Eq} \\
&&\Sigma_{t t'} = -\frac{1}{2} \sqrt{\frac{6\epsilon}{53}} 
  (\gamma(t-t'))^2 \quad, \quad \sigma(t) = - \int_{t_i}^t d {t_1} 
\Sigma_{tt_1} \label{kernel_sigma}
\end{eqnarray}
where $\gamma(x) = (x+\Lambda_0^{-2})^{-1}$, $\Lambda_0$ being the UV
cutoff, and $\tilde \zeta(t)$ is an effective 
disorder induced Gaussian noise with zero mean and correlations
$\langle \tilde \zeta(t)\tilde \zeta(t')\rangle_{{eff}}  = 2
T \delta(t-t') 
+ D_{tt'}$ with:
\begin{eqnarray}
D_{t t'} = \frac{T_c}{2}
  \sqrt{\frac{6\epsilon}{53}} 
  \left( \gamma(t-t') - \gamma(t+t')\right) \label{kernel_d}
\end{eqnarray}
The idea is then to compute 
$\overline{{P}_c}(t)$ as the persistence probability of the
process $\tilde 
M(t)$. The rhs of Eq.~(\ref{Eff_Eq}) clearly indicates
that this process is non-Markovian. However, as $\Sigma_{tt'}$
(\ref{kernel_sigma}) as well 
as $D_{tt'}$ (\ref{kernel_d}) are of order 
${\cal  O}({\sqrt{\epsilon}})$, one can use the perturbative computation of
$\theta_c$ around a Markov process initially developed in
Ref.~\cite{satya_clement_persist} and further studied in the context
of critical dynamics in Ref.~\cite{oerding_persist}. In that
purpose~\cite{satya_review}, let us
introduce the normalized Gaussian process ${m}(t) = {\tilde
  M(t)}/\sqrt{{\langle \tilde M^2(t)\rangle_{\mathrm{eff}}}}$. Let 
$T=\ln{t}$, then ${m}(T)$ is a {\it stationary} Gaussian process and its
persistence properties are obtained from the autocorrelation function:
\begin{equation}
\langle{m}(T) {m}(T_w)\rangle_{\mathrm{eff}} = e^{-\mu(T-T_w)}{\cal
      A}(e^{T-T_w}) \quad, \quad {\cal A}(x) = \left[1 +
  \frac{1}{4}\sqrt{\frac{6\epsilon}{53}}\left(x \log{\frac{x-1}{x+1}} -
  \log{\frac{x^2-1}{4 x^2}}    \right)  \right]          \label{TTI} 
\end{equation}     
with $\mu ={(\lambda_c-d+1-\eta/2)}/{z}$.
Under this form (\ref{TTI}), one can use the first order perturbation
theory result of Ref.~\cite{oerding_persist} to obtain the one-loop
estimate:   
\begin{eqnarray}
\Delta \equiv  \theta_c - \mu = \sqrt{\frac{6
    \epsilon}{53}} \frac{\sqrt{2}-1}{2} 
+ {\cal O}(\epsilon) = 0.06968... \quad {in} \quad d=3,
    \label{persist_one_loop} 
\end{eqnarray}
where $\mu$ is the value corresponding to a Markov
process. The second term in Eq.~(\ref{persist_one_loop}) is the first
correction due to the non-Markovian
nature of the dynamics. Interestingly, this correction is entirely
determined by the non-trivial structure of the scaling function ${\cal A}(x)$,
which is directly obtained from $F_{\cal C}(v,u)$
(\ref{janssenscalingcorr}). Notice 
also that, at variance with 
the pure ${O}(N)$ models 
\cite{majumdar_critical, oerding_persist}, these corrections in
presence of quenched static disorder already
appear at one-loop order.

\subsection{Numerical simulation in $d=3$}

We now turn to the results from our Monte Carlo simulations of the
relaxational dynamics of the randomly diluted Ising model
(\ref{def_diluted}) in dimension $d=3$, which were performed on $L^3$ 
cubic lattices with periodic boundary conditions. The system 
is initially prepared in a random initial configuration with zero mean
magnetization $M_0 = 0$. Up and down spins are randomly 
distributed on the occupied sites, mimicking a high-temperature
disordered configuration before the quench. At each time step, one
site is randomly chosen and the move  
$s_i \to - s_i$ is accepted or
rejected according to Metropolis rule. One time unit corresponds to
$L^3$ such time steps. The exponent $\theta_c$ is measured numerically
for cubic lattices of linear 
size $L =$ 8, 16, 32 and 64. After a quench to $T_c$ from the initial
random configuration each system evolves until the global
magnetization first change sign. $\overline{P}_c(t)$ is then measured
as the fraction of surviving systems at each time $t$, over a
number of samples which 
varies from $2 \times 10^5$ for $L=8$ to $2 \times 10^4$ for
$L=64$. 
\begin{figure}
\begin{minipage}[b]{0.43\linewidth}
\includegraphics[width=18pc]{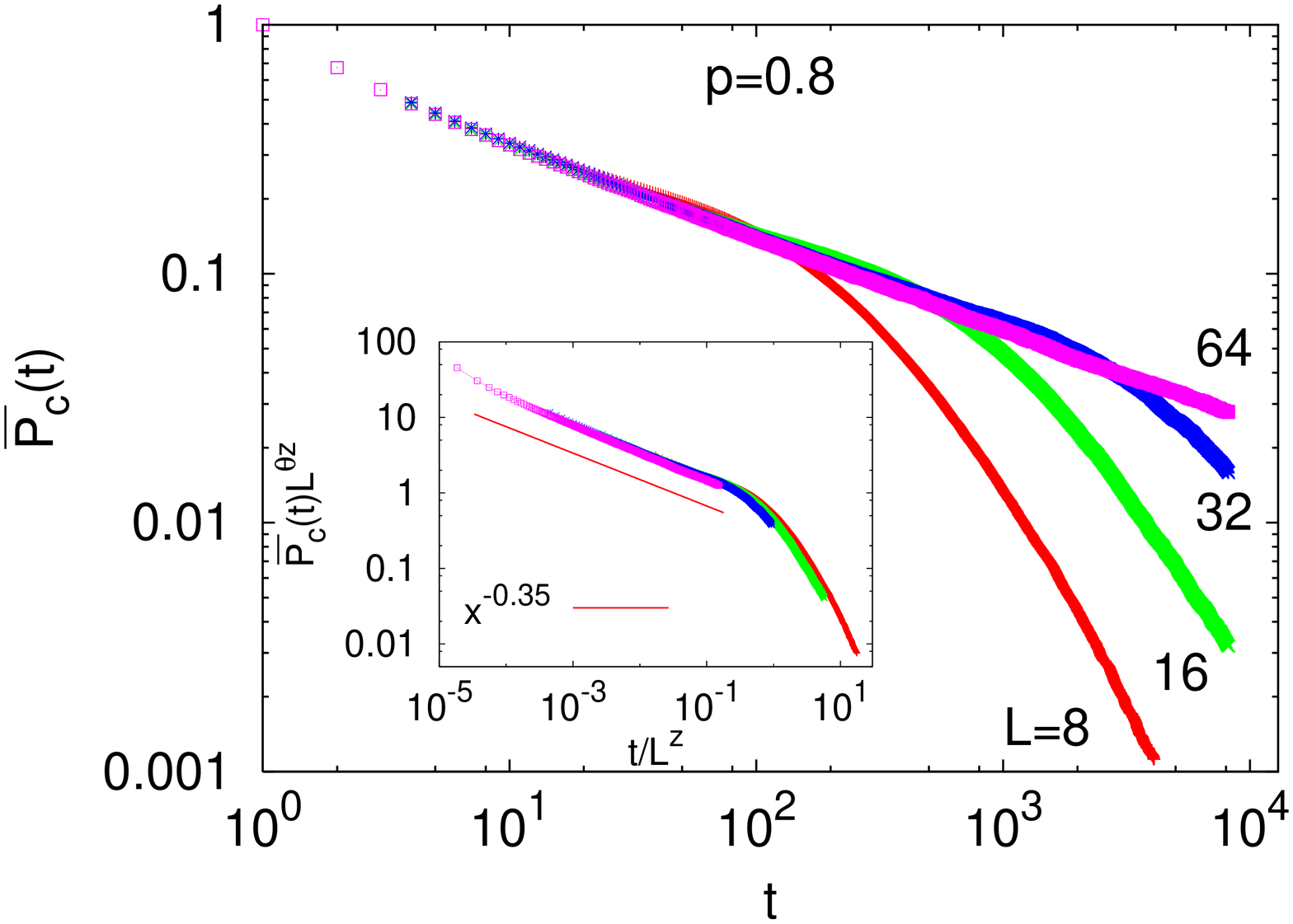}
\caption{Persistence probability $\overline{P_c}(t)$ plotted in the
  log-log scale for $p=0.8$ and different $L$, with $M_0=0$. {\bf
    Inset:}  
  $L^{\theta_c z} \overline{P_c}(t)$ vs $t/L^z$ for different $L$
with $z=2.62$ and $\theta = 0.35$.}
\label{fig_persist1}
\end{minipage}\hspace{2pc}%
\begin{minipage}[b]{0.43\linewidth}
\vspace*{-2cm}
%\begin{figure}
\includegraphics[width=18pc]{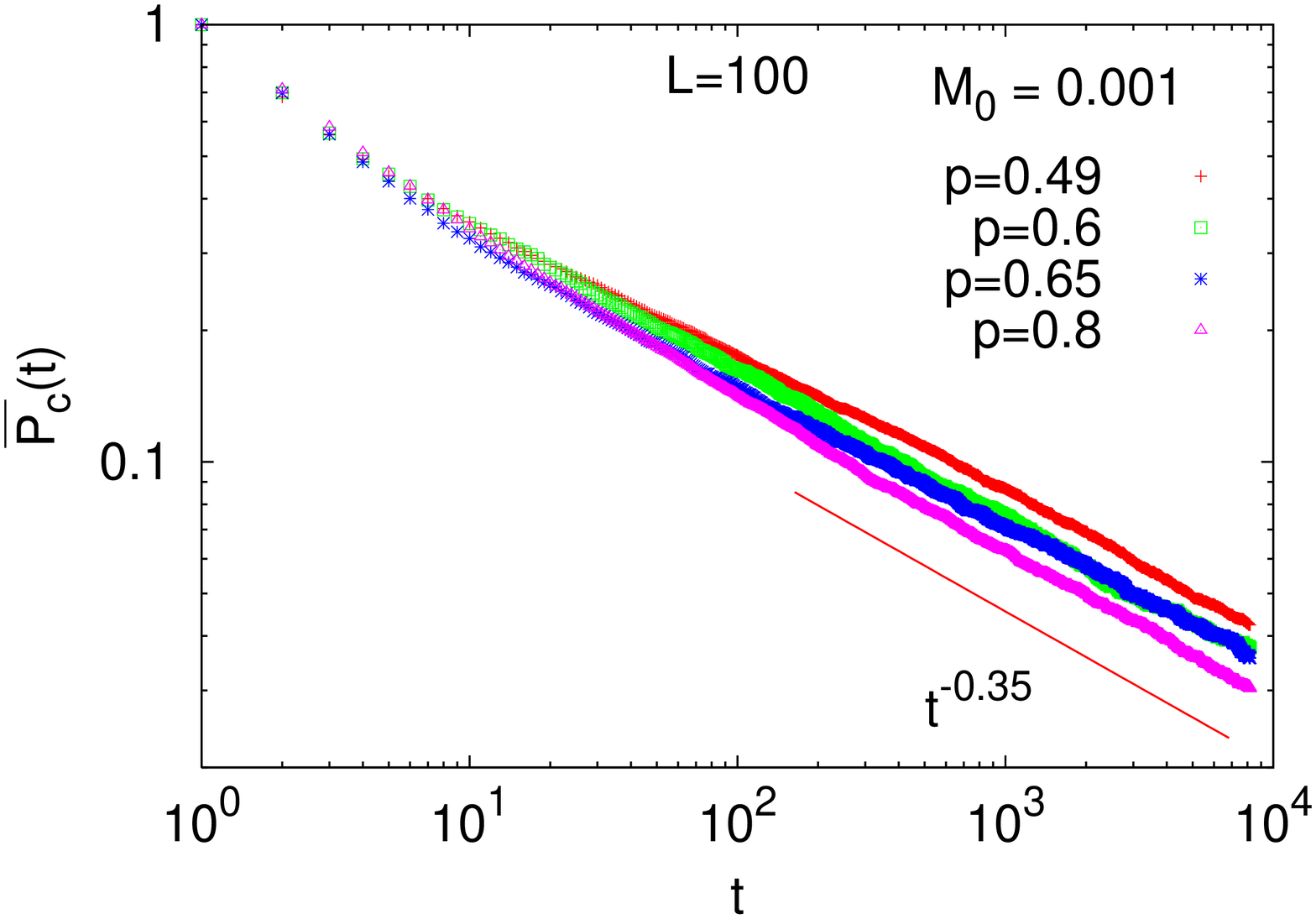} 
\caption{Similar to
  Fig.~\ref{fig_persist1}, but prepared with a non-zero 
  initial magnetization $M_0=0.001$. The linear system size in this
  case is $L=100$.
\\
}\label{fig_persist2} 
%\end{figure}
\end{minipage}
\end{figure}

In Fig.~\ref{fig_persist1} we present the results of
${\overline P}_c(t)$ for 
$p=0.8$ and for 
different lattice sizes. According to standard finite-size
scaling~\cite{majumdar_critical}, one expects the scaling form
$\overline{P}_c(t) = t^{-\theta_c} f(t/L^z)$,
where $z$ is the dynamical exponent. Keeping the rather well
established value of
$z=2.62(7)$~\cite{parisi_simu_rim, us_pre} fixed, $\theta_c$
is varied to 
obtain the best data collapse. The final scaled plot is shown in
the inset of  
Fig.~\ref{fig_persist1}. This allows for a first estimate of the
exponent $\theta_c$
\begin{eqnarray}\label{num_value}
\theta_c = 0.35 \pm 0.01
\end{eqnarray}

We have also computed the persistence probability for systems quenched from 
random configuration with a small initial magnetization
$M_0=0.001$. The number of up $N_{\mathrm{up}}$ and down
$N_{\mathrm{down}}$ spins are 
thus : $N_{\mathrm{up}} = (1+M_0)/{2N_{\mathrm{occ}}}$ and $N_{\mathrm{down}} =
N_{\mathrm{occ}}-N_{\mathrm{up}}$. First we randomly distribute the
$N_{\mathrm{up}}$ up spins in 
the occupied sites of the lattice and then fill up the rest with down spins.
As noticed previously~\cite{Schulke97} this protocol
allows to reduce the statistical noise and thus to study larger system
sizes (this however renders the finite size scaling analysis more
subtle~\cite{Schulke97}). In Fig.~\ref{fig_persist2}, we plot
the persistence 
for system size $L=100$ ( the data have been averaged over $2\times 10^4$
ensembles) for $p=$ 0.499, 0.6, 0.65 and 0.8. The study of larger
system size allows to reduce the corrections to scaling. Indeed, 
although in the short time scales, the
straight lines have slightly different slopes which depends upon
$p$, at later times the slopes varies from $0.36(1)$ for
$p=0.8$ to $0.35(1)$ for $p=0.499$ :  this confirms the
$p$-independent value of $\theta_c$ obtained previously
Eq.~(\ref{num_value}). 

In order to compare this numerical value (\ref{num_value}) with our
one-loop calculation (\ref{persist_one_loop}) one needs an estimate
for $\mu$. Such an estimate is needed
not only for the sake of this comparison but also to characterize
quantitatively non Markovian effects. The argument mentioned in
the introduction, relying on the CLT, which says
that the global magnetization is, in the thermodynamic limit, a
Gaussian variable is also valid in the presence of disorder, and we
have checked it numerically. Therefore, a
finite difference $\Delta$ (\ref{persist_one_loop}) is the signature
of a Non Markovian 
process. Because of the relatively big
error bars on $\lambda_c$ and $z$, we propose, 
alternatively, to express $\mu$ in terms of the initial slip exponent
, $\theta'$ 
\cite{janssen_rg}, using
$\lambda_c= d-z\theta'$, as $\mu = -\theta' + (1-\eta/2)/z$. Using our
previous estimate for the initial slip exponent $\theta' = 0.10(2)$
(\ref{initial_slip_num}), this gives  
$\mu_{\mathrm{num}} = 0.27(3)$ and our numerical estimate 
\begin{eqnarray}
\Delta_{\mathrm{num}} = 0.08 \pm 0.04 
\label{estim_delta}
\end{eqnarray} 
which is in good agreement with our previous one-loop estimate
(\ref{persist_one_loop}). We 
also notice that these deviations from a Markov process are
slightly larger than for the pure case~\cite{Schulke97}. 

\section{Conclusion}

In conclusion, we have studied, analytically using the Exact
renormalization group equation as well as numerically, different
aspects of non-equilibrium dynamics at a random critical
point. Interestingly, concerning the violation of FDT, although the
heuristic argument of 
Ref. \cite{calabrese_on_oneloop} does not hold for the present model,
we have shown explicitly that the limiting FDR for the local and
global magnetization do coincide (\ref{expl_calc}). And it would be
interesting to compare 
this perturbative calculation with numerical simulations. In addition,
in view of recent studies
\cite{us_pre,ci,berthier_xy,zheng_review,fedorenko_ci}, it would be  
also interesting to study this FDR for the 
relaxational dynamics following a quench from a completely ordered
initial condition. Finally,
concerning the persistence properties, we have shown that the RG
calculation, together with the perturbative methods of
Ref. \cite{satya_clement_persist, oerding_persist}
allows for a rather precise estimate of $\theta_c$. And it would be
interesting to extend this kind of approach deep inside the glassy
phase of disordered systems.

\ack
GS acknowledges the financial support provided
through the European Community's Human Potential Program 
under contract HPRN-CT-2002-00307, DYGLAGEMEM.\\

\end{document}